# Hydroxide Transport and Mechanical Properties of Polyolefin-Based Anion Exchange Membranes from Atomistic Molecular Dynamics Simulations

Mohammed Al Otmi,[1] Ping Lin,[2] William Schertzer,[3] Coray M. Colina,[2,3,4] Rampi Ramprasad,[3] Janani Sampath[1,*]

[1]Department of Chemical Engineering, University of Florida, Gainesville, Florida, 32611

[2]Department of Chemistry, University of Florida, Gainesville, Florida, 32611

[3]Department of Materials Science and Engineering, University of Florida, Gainesville, Florida, 32611

[4]George and Josephine Butler Polymer Research Laboratory, University of Florida, Gainesville, Florida, 32611

[5]School of Materials Science and Engineering, Georgia Institute of Technology, Atlanta, Georgia, 30332

[*]jsampath@ufl.edu

**Abstract**

Anion exchange membranes are used in alkaline fuel cells and offer a promising alternative to the more expensive proton exchange membrane fuel cells. However, hydroxide ion conductivity in anion exchange membranes is low, and the quest for membranes with superior ion conductivity, mechanical robustness, and chemical stability is ongoing. In this study, we use classical molecular dynamics simulations to study hydroxide ion transport and mechanical properties of eight different hydrated polyolefin-based membranes, to provide a molecular-level understanding of the structure-function relationships in these systems. We examine the microstructure of the membranes and find that polymers with narrow cavity size distribution have tighter packing of water molecules around hydroxide ions. We estimate the self-diffusion coefficient of water and hydroxide ions and find that water molecules have a higher diffusion than hydroxide ions across all systems. The trends in hydroxide diffusion align well with experimental conductivity measurements. Water facilitates hydroxide diffusion, and this is clearly observed when the hydration level is varied for the same polymer chemistry. In systems with narrow cavities and tightly bound hydroxide ions, hydroxide diffusion is the lowest, underscoring the fact that water channels facilitate hydroxide transport. Finally, we apply uniaxial deformation to calculate the mechanical properties of these systems and find that polymers with higher hydration levels show poor mechanical properties. Atomistic molecular dynamics models can accurately capture the trade-off between hydroxide transport and mechanical performance in anion exchange membranes and allow us to screen new candidates more efficiently.

**Introduction**

Polymer anion exchange membranes (AEMs) which contain a positively charged functional group and facilitate the transport of hydroxide anions are widely studied, as they are used in alkaline fuel cells. AEMs offer a less expensive alternative to proton exchange membranes (PEMs) such as Nafion which have been extensively used due to their stability, processability, and high conductivity.[1] While PEM fuel cells exhibit excellent performance, their reliance on costly catalysts, notably platinum, has posed economic challenges.[2] In contrast, alkaline fuel cells have faster reaction kinetics, which facilitate the utilization of less expensive catalysts.[3] This has sparked considerable interest in both academic and industrial sectors towards identifying optimal AEMs to be used in alkaline fuel cells.

While AEMs demonstrate remarkable energy conversion efficiency, the quest for an AEM material with superior ion conductivity, robust mechanical properties, and high chemical stability is ongoing.[4] The diffusion of hydroxide ions ($OH^-$) in AEMs is four times slower compared to the diffusion of protons ($H^+$) in PEMs. To achieve conductivity comparable to PEMs, a higher ion exchange capacity (IEC) is required.[5] However, a high IEC usually leads to considerable swelling of the membrane, which compromises the mechanical robustness of the membrane module.[6] Furthermore, quaternary ammonium functional groups which constitute the polymer cations in many AEMs, undergo nucleophilic attack by hydroxide ions leading to membrane degradation.[7] Several studies have focused on modifying the side chain of quaternary ammonium functional group to improve both conductivity and chemical and mechanical stability.[8–10]

To enhance conductivity without compromising the mechanical and chemical stability, we need a clear understanding of hydroxide ion transport through the polymer matrix. Key contributors to ion transport include morphology of polymer membranes, and water uptake. These factors also control the mechanical integrity of the membrane, making it imperative to study the morphology (or microstructure) of the polymer as a function of water uptake to improve emergent properties, such as ion transport and membrane strength. When hydrated, pores in the polymer matrix undergo swelling resulting in the creation of continuous water channels that facilitate both vehicular (convective transfer) and Grotthuss (proton hopping) diffusion of hydroxide ions.[11] At low hydration levels, hydroxide ions neutralize the cationic sites, degrading the membrane.[12] Therefore, moderate to high hydration levels are maintained in AEMs, and ion transport is mediated primarily by water.[13,14] Even then, OH⁻ dynamics are challenging to characterize as the motion of the hydroxide ions is coupled with the polymer backbone, water molecules, as well as other OH⁻ ions. A recent study used quasi-elastic neutron scattering (QENS) to decouple water and polymer relaxation dynamics from the diffusional dynamics of hydroxide ions.[15] The authors suggest that the diminished efficiency of anion transport at lower hydration levels arises from the lack of coupling with water diffusional dynamics. Water molecules are needed to facilitate OH⁻ transport, and a molecular-level understanding of the OH⁻ and water transport mechanisms within AEMs is critical for optimizing their performance and efficiency.

Molecular dynamics (MD) simulations can provide important insights into the molecular level interactions and the transport mechanism of hydroxide ions in AEMs. By employing detailed ab-initio calculations, Zelvich et al. demonstrated that OH⁻ ions attract a cluster of water molecules, with the number of molecules in each cluster ranging from one to five, depending on the hydration level.[16] The first solvation shell forms at a distance of 3.7 Å from the OH ion, encompassing 1-3 water molecules depending on the hydration number. In the bulk solution, the first hydration shell around OH⁻ comprises four water molecules.[16–18] ReaxFF molecular dynamics simulations used radial distribution function to show that the first hydration shell contains 3.5 and 4.2 water molecules in low and high hydration levels, respectively.[19] The contribution of vehicular and Grotthuss diffusion is dependent on the hydration number ($\lambda$); at $\lambda = 4$ or lower, water mediated $OH^-$ hopping (Grotthuss) is the dominant mechanism.[13,15] The study by Chen et al. also showed that a combination of both vehicular and Grotthuss diffusion contribute to the overall transport, but emphasized that the vehicular diffusion is significantly more dominant at hydration levels of $\lambda > 14$.[20] Zhang and van Duin investigated three functionalized poly (phenylene oxide) polymers using both classical and reactive forcefields. The study examined the effect of water content and showed that $OH^-$ diffusion increases with the increase in the hydration level. They also found that increasing the alkyl chain length on the cationic side of quaternary ammonium protect $N^+$ from nucleophilic attack.[19] Another study delved into $OH^-$ transport mechanism using polarizable forcefield, and highlighted the effect of bottlenecks formed in the polymer matrix.[5] The study shows that the formation of narrow regions (bottlenecks) influences the water channels, and has an adverse effect on ion transport. Transport through these narrow bottlenecks via vehicular diffusion is thermodynamically unfavorable as it requires the hydroxide ion to partially dehydrate first leading to a significantly higher contribution from Grotthuss transport. Limiting the number of bottlenecks and achieving a desirable water channel morphology can lower the energy barrier for hydroxide transport through vehicular diffusion.[5] Other studies using both classical and reactive force fields have shown that the Grotthuss diffusion dominates in confined environments, and have highlighted the importance of Grotthuss hopping when dealing with sub-nanometer wide water channels. [21,22]

In this study, we focus on how polymer morphology affects convective $OH^-$ transport and mechanical properties, in polyolefin based AEMs by employing classical molecular dynamics (MD) simulations. We systematically analyze eight systems that have been characterized experimentally and offer molecular insights into experimentally observed trends by relating ion dynamics to polymer chemistry. These systems were chosen as they have the same backbone structure, allowing us to correlate functional group chemistry to overall performance. We consider high hydration levels where vehicular (convective) transport is the primary contributor to hydroxide dynamics. We investigate the mechanical properties of these AEMs, as this is important for ensuring the integrity and longevity of the membranes under different operating conditions. Understanding how ions move in the polymer matrix and how water channels and the structure of the polymer influence ion transport is crucial for optimizing AEM design and enhancing their properties. We expect that these insights can guide the design of novel AEMs with enhanced $OH^-$ dynamics and mechanical properties.

**Materials and Methods**

**System Setup**

Three distinct polyolefin systems featuring similar backbone chains are selected to investigate how side chain architecture impact the performance and stability of AEMs (Figure 1). These polymers chosen were also extensively probed via experiments. Specifically, the chosen polymers are ammonium-functionalized polyethylene (PE-F) that contain two different cationic functional groups, with different alkyl chain branching, as detailed in the study by Wang et al.[23]; tetraalkylammonium-functionalized polyethylene (TFP-F), as studied by Kostalik et al [24]; and ammonium functionalized polypropylene (PP-F) with two functional groups from the work of Zhang et al. [25] For simplicity, we will use the abbreviations PE-F, TFP-F, and PP-F to denote the three systems. These abbreviations which we will use throughout the manuscript, correspond to the functionalized polymers depicted in Figure 1.

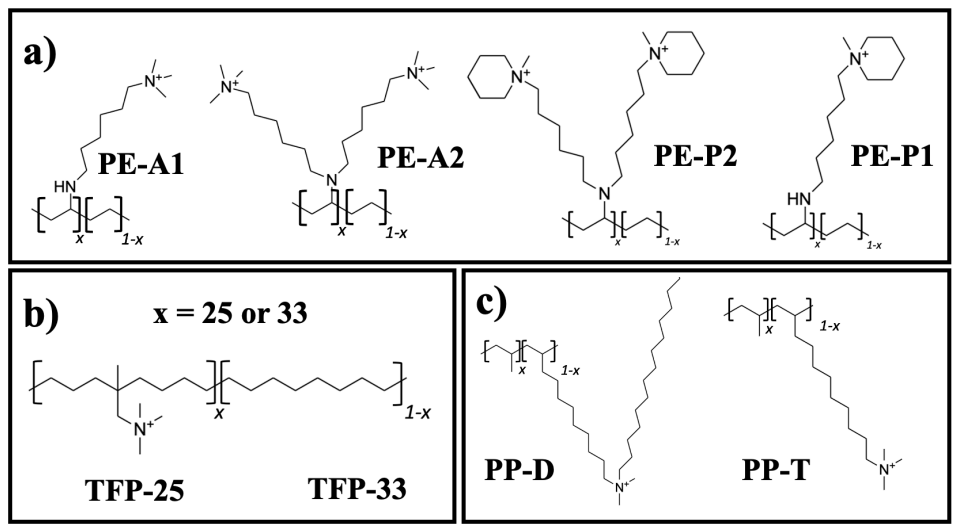

**Figure 1: Anion exchange membrane chemistries investigated in this study. Eight polyolefins are divided into three sets of systems depending on the backbone architecture and functional groups: (a) ammonium-functionalized polyethylene (PE-F) (b) tetraalkylammonium-functionalized polyethylene (TFP) (c) Ammonium-functionalized polypropylene (PP-F).**

For the first set of ammonium-functionalized PE, we vary the number of alkyl chains attached to the polymer backbone (1 or 2), as well as the type of functional group at the end of the alkyl chains. The alkyl chain length is constant, with 6 -$CH_2$ groups. The backbone is 13% functionalized with either Trimethylammonium (TMA) or N-methylpiperidinium (NMP) groups. These particular compositions are selected based on their superior performance, as documented in the study by Wang et al.[23] Systems are referred to as PE-A1 and PE-A2 if they are functionalized with 1 or 2 TMA groups, respectively, and PE-P1 and PE-P2, if they contain 1 or 2 NMP groups, respectively. For the second family of polymers (TFP), ion contents of 29% and 33% were examined experimentally. To remain as close to prior study[24], we consider ion contents of 25% and 33%, and these systems are designated as TFP-25 and TFP-33, respectively (Figure 1). We opted to implement larger difference in ion content than that used in experiments so that the difference in performance as a function of ion concentration is clear in the simulations. A maximum ion content of 33% is utilized to prevent undesirable membrane module swelling, as exceeding this threshold leads to significant expansion as observed by Kostalik.[24] In the third family, ammonium functionalized polypropylene (PP-F), we vary the length of the alkyl group attached to trimethylammonium to investigate the steric effects introduced by the extended alkyl chain. The side chain on the backbone is terminated with trimethylammonium and N,N-dimethyl-1-hexadecyamine, and the polymers are referred to as PP-T and PP-D, respectively. The presence of a lengthy alkyl chain in PP-D acts as a spacer between distinct cationic groups, mitigating nucleophilic attacks on these cationic groups and enhancing the membrane's stability.[25]

Monomer units are constructed using Avogadro[26], followed by geometry optimization using the semi-empirical EFN2-xTB method[27]. The assignment of the force field parameters is carried out using Antechamber.[28] Both atom types and force field parameters are determined in accordance with the General Amber Force Field (GAFF2).[29] The AM1-BCC method is used to assign atomic partial charges.[30] Explicit water molecules are inserted, and their interactions are represented using a flexible SPC/E water model.[31] Subsequently, Pysimm[32] is employed to generate various polymer chains using the random walk simulated polymerization algorithm. Ions and water molecules are incorporated into the system using Packmol.[33] The number of $OH^-$ ions is varied to investigate different ion exchange capacities, while adjustments are made to hydration levels to replicate the reported experimental water uptake. In all systems, water plays a crucial role in enhancing ion conductivity, necessitating an adequate water uptake to establish interconnected hydrated domains that facilitate ion diffusion.[34–37] In our simulations, we used the reported water uptake values to calculate the hydration number as defined in equation 1. Here, the hydration number ($\lambda$) is determined using the theoretical ion exchange capacity (IEC, mmol[OH]/g), the molecular weight of water ($M_{wt}$, g/mol), the experimental water uptake (WU% $H_2O$), and a constant factor 10 for unit adjustment. Theoretical IEC is determined from the elemental analysis of AEM composition. Four independent replica simulations with different starting configurations (which includes independent monomer packing and polymerization) are carried out for each of the eight systems described above, for a total of 32 systems. The system parameters are listed in Table 1.

$$\lambda = \frac{WU\% \ H_2O * 10}{IEC * M_{wt}} \qquad (1)$$

**Table 1: System details for the eight polymers considered. The box length is averaged over four replicas.**

| System | Monomers, Chains | Box Length (nm) | Theoretical IEC (mmol OH/g) | Water Uptake % | # OH ions | Hydration # ($\lambda$) |
|---|---|---|---|---|---|---|
| **PE-A1** | 460, 4 | 6.5 | 2.3 | 75 | 240 | 18 |
| **PE-A2** | 460, 4 | 7.7 | 3.1 | 104 | 480 | 18 |
| **PE-P1** | 460, 4 | 6.1 | 2.1 | 30 | 240 | 8 |
| **PE-P2** | 460, 4 | 7.3 | 2.8 | 56 | 480 | 11 |
| **TFP-25** | 320, 10 | 5.3 | 1.8 | 97 | 80 | 30 |
| **TFP-33** | 300, 10 | 5.5 | 2.3 | 132 | 100 | 32 |
| **PP-T** | 250, 4 | 5.6 | 2.5 | 34 | 200 | 7 |
| **PP-D** | 250, 4 | 6.6 | 1.6 | 37 | 200 | 12 |

After the replicas are constructed, we employ a 21-step annealing protocol[38] to achieve system equilibration, during which each replica is allowed to relax to its equilibrium density. Then all replicas undergo an additional 1 ns run in the NVT ensemble, at a temperature of 300 K. Following this, the replicas continue to run in the same ensemble for 20 ns (hydroxide ions and water molecules attain diffusive behavior over this period), during which data is collected. Data shown is the average over the four replicas for each system. All simulations in this work utilized Large Scale Atomic Massively Parallel Simulator (LAMMPS).[39] Nosé-Hoover thermostat with a damping factor of 100 ps is used in the equilibration and production simulations, and Nosé-Hoover barostat is coupled to the thermostat with a pressure damping parameter of 100 ps during the 21-step annealing process. A velocity-verlet integrator with a 1 femtosecond timestep is employed. Long ranged interactions are accounted for by employing a particle-particle/particle-mesh (PPPM) algorithm with 1.5 nm cutoff.

**Analysis**

To assess water channels across the different hydrated polymer samples, we examine the channel dimensions, or cavities within the polymer matrix. This entails removal of ions and water molecules from different samples and characterizing unoccupied regions of the polymer matrix (which would be occupied by ions and water in the hydrated polymer). To accomplish this, we take a fully saturated membrane at the end of the 20 ns run, remove all ions and water molecules from this membrane, and quantify the voids which are left behind in place of the water molecules and ions. Then, we apply Void Analysis Codes and Unix Utilities for Molecular Modeling and Simulations, or VACUUMMS.[40,41] VACUUMMS is an open-source software package that leverages the Cavity Energetic Sizing Algorithm (CESA)[42], which is a Monte Carlo-based energetic technique employed to characterize voids within a polymer matrix. In the CESA method, cavities are defined as spherical volumes with energy centers, which correspond to local minima in the repulsive energy field of the particles. This approach allows for the determination of a size distribution of voids, collectively constituting the hydration channels within the material. We have used this method in the past to quantify voids in gas separation membranes.[43]

To quantify the local packing of hydroxide ions near water molecules and the polymer cation, we utilize the radial distribution function (RDF), $g_{ij}(r)$, which is the normalized probability of an atom of type $i$ existing at a distance r from an atom of type $j$. Specifically, we calculate $g_{H^*-H}(r)$, $g_{O^*-O}(r)$, $g_{O^*-N}(r)$ and $g_{N-N}(r)$, where H* and O* are the hydroxide hydrogen and oxygen, and H and O are water hydrogen and oxygen, respectively, and N is the repeating nitrogen cation on the polymer backbone. We use the radial pair distribution function extension in VMD to compute $g_{ij}(r)$, and we average over the 20 ns runs of all replicas to compute each RDF.

To capture the dynamics of the hydroxide ions, water molecules, and polymer chains, we compute the mean squared displacement (MSD) for each species, given by:

$$MSD(t) = <(r_i(t) - r_i(0))^2) \qquad (2)$$

Here, $r_i(0)$ refers to the initial frame after 1 ns, and we do not perform any block averaging. Each MSD curve is averaged over four independent replicas. For each component, (polymer chains, hydroxide ions, water), we compute MSD and plot it against time in a logarithmic scale. Polymer MSD is averaged over all atoms of the polymer. Once the MSD of OH⁻ and water attain diffusive behavior in each system (characterized by long time MSD slope ~ 1), the diffusion coefficient can be calculated using Einstein's equation:

$$MSD(t) = 6Dt \qquad (3)$$

Where D is the diffusion coefficient, and MSD is the value of the mean squared displacement at a particular time t within the diffusive regime. In the simulation times considered, both water and $OH^-$ ions reach the diffusive regime (long time MSD slope ~ 1), allowing us to calculate their diffusion coefficient from Equation 3.

To assess the mechanical properties of the hydrated polymers, each equilibrated system is subjected to a uniaxial tensile deformation along the x-axis with a constant engineering strain rate, by increasing the box dimension in finite steps. The barostat is applied in the y and z directions, to maintain an overall pressure of 1 atm. Employing the *erate* style in LAMMPS, we apply a strain rate of $10^{-5}$ $s^{-1}$ for a duration of 200,000 fs, reaching a strain of 2.0 (equivalent to three times the initial box length). Deformation is performed across all four replicas. We refer the reader to prior work for further details about the protocol.[44]

**Results and Discussion**

**Structure**

The two primary modes of hydroxide transport in AEMs are vehicular diffusion and Grotthuss hopping. The microstructure of the polymer, which is directly responsible for the degree of water uptake, controls the ratio of vehicular vs. hopping mechanism in a given system. Thus, it is important to characterize the microstructure of the polymer and map the hydration channels. We do this by employing the cavity energetic sizing algorithm VACUUMMS, as described in the Methods section, which gives the pore size distribution (PSD) for each system. We characterize water channels qualitatively, through visual representations in the form of snapshots of one-nanometer cross-section of various simulation boxes (Supplementary Information).

Figure 1 shows the PSD in the PE family of AEMs. We find that the cavity size distribution is in the order PE-A2 > PE-A1 > PE-P2 > PE-P1. Overall, PE-A1 and PE-A2 have large cavities that are > 1 nm in diameter, and PE-P1 and PE-P2 have small cavities that are ~0.2 nm in diameter. PE-A2, with the double functional TMA groups (Figure 1c) has the largest cavities and the broadest cavity distribution. This is primarily due to steric effects, due to the inability of the bulky TMA group to pack efficiently, giving rise to large water channels. On the other end of the spectrum, we have PE-P1 with the smallest cavity size. Previous studies have shown that polymers with cyclic groups pack compactly, due to pi-pi stacking.[45] This could be why PE-P1 and PE-P2, with the aromatic functional groups, are able to pack more efficiently, resulting in fewer cavities and smaller water channels that are smaller in size compared to PE-A1 and PE-A2, which contain TMA functional groups. Both PE-A1 and PE-A2 have a large cavity distribution and larger water channels compared to PE-P1 and PE-P2 due to the bulky TMP functional group, leading to a more porous microstructure. The snapshots of the water channels are also in agreement with the pore size distribution, with PE-P1 having sparsely connected narrow channels, and PE-A2 has dense water channels (Figure S2, Supporting Information). From the snapshots, we see that the water molecules surround the cationic groups along the polymer backbone, and the OH$^-$ ions are associated with the cations as well.

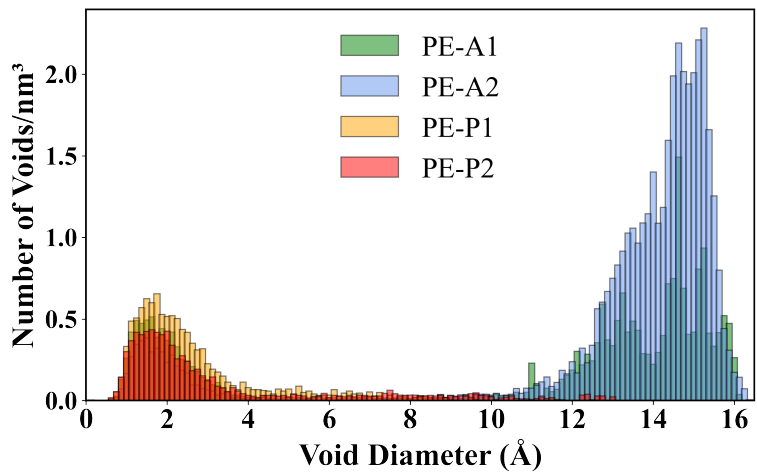

**Figure 2: Cavity size distribution showing the number of cavities normalized by the box volume as a function of cavity diameter, calculated using VACUUMMS for four PE polymers, as labeled.**

The cavity distribution and water channels for the TFP polymers are shown in Figure 3. The voids in both TFP-25 and TFP-33 have a bimodal distribution, with narrow (<0.4 nm) and large (>1.2 nm) voids. There is not a significant difference in the void distribution between the two systems, which is to be expected given that the polymers have the same backbone and sidechain architecture. We do find that the cavity distribution for the 25% system shows a higher number of both narrow and broad voids. The challenge in distinguishing differences in the polymer microporosity between the two TFP polymers arises from the substantially elevated and closely matched hydration levels ($\lambda$= 30, 32), along with very similar chain architecture. However, upon examining the y-axis in Figure 2 and Figure 3, the TFP polymers have higher number of voids compared to the polymers in the PE systems. This is also observed qualitatively in the snapshots (Figure S2, S3 in Supporting Information).

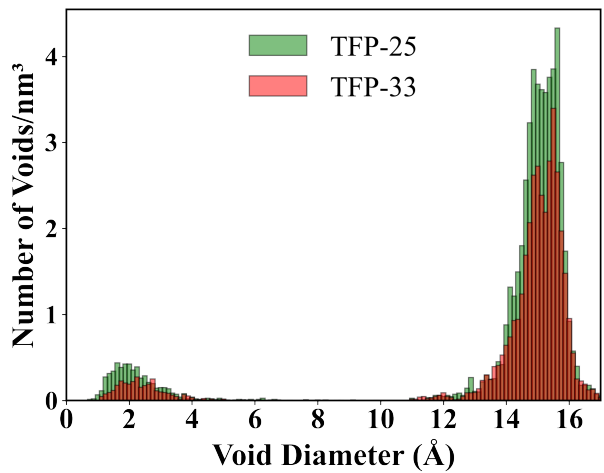

**Figure 3: Cavity size distribution showing the number of cavities normalized by the box volume as a function of cavity diameter, calculated using VACUUMMS for TFP-25 and TFP-33 polymers.**

In the ammonium functionalized PP polymers, PP-D with the larger alkyl spacer has higher number of cavities and a broader distribution, compared to PP-T which consists of significantly fewer cavities. PP-D has cavities that are > 1.0 nm in diameter, whereas all the cavities in PP-T are less than 1 nm. Specifically, PP-D has a long alkyl chain in place of a methyl group, and the presence of this chain acts as a spacer between the cationic groups of the polymer, which enhances microphase separation and promotes the formation of larger water channels. This is why the water channel is significantly larger in PP-D compared to PP-T. This distinction is evident in the distribution depicted in Figure 4 and the snapshot in Figure S4 of the Supporting Information. PP polymers have the fewest cavities in a given volume, signifying that this family of polymers has the lowest cavities and smallest water channels.

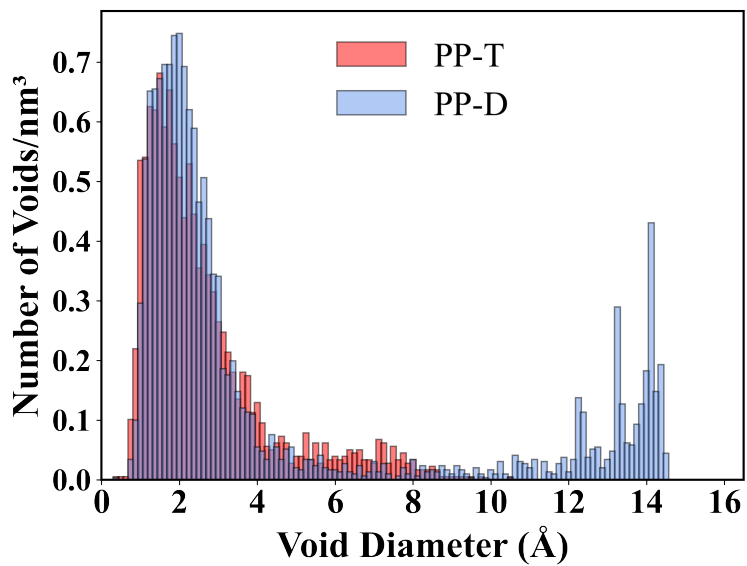

**Figure 4 Cavity size distribution showing the number of cavities normalized by the box volume as a function of cavity diameter, calculated using VACUUMMS for PP-T and PP-T polymers.**

Next, we calculate the packing of hydroxide ions near the polymer and water, as well as polymer-polymer packing by computing the radial distribution function (RDF), as described in the Methods section. Both the distribution of OH⁻ ions within the polymer matrix and the spatial arrangement of water molecules surrounding these OH⁻ ions are critical parameters that significantly influence ion transport and thus, conductivity in AEMs. The $H^*$–H (hydroxide hydrogen – water hydrogen) and the $O^*$–O (hydroxide oxygen – water oxygen) RDF show the arrangement of water around hydroxide ions. The first peak in $O^*$–O and the first two peaks in $H^*$–H correspond to the first coordination shell, or the closest water molecules around OH⁻. Across all eight systems, the first peak in $O^*$–O has a high intensity, as the hydroxyl group forms a strong hydrogen bond with the water oxygen, causing both hydroxyl hydrogen and oxygen to pack closely to water molecules.

Figure 5 shows RDFs for the PE polymers. PE-A1 and PE-A2 have the same hydration number of 18, implying that differences in the local packing is directly correlated to the nature of the functional groups. We find that the water – hydroxyl ($g_{H^*-H}(r)$ and $g_{O^*-O}(r)$) peaks are sharpest in PE-P1 system (Figures 5a and 5b), suggesting that the water molecules in PE–P1 are closer and more tightly ordered around the hydroxyl ions than the other PE polymers. Additionally, the OH⁻ ions in PE-P1 are packed closer to the cation on the polymer backbone, compared to the other polymers (Figure 5c). Conversely, PE-A2 shows the smallest $g_{H^*-H}(r)$, $g_{O^*-O}(r)$, and $g_{O^*-N}(r)$ peaks. As for the other two systems, (PE-P2 and PE-A1), their water – hydroxyl and polymer – hydroxyl peak intensities are in between these extremes. These results indicate that in polymers with narrow cavity size distribution (Figure 2), the hydroxyl ions are packed closer to both water molecules and cationic groups along the polymer sidechain. We also find that the polymer – polymer packing, $g_{N-N}(r)$ (Figure 5d) is the highest in PE-P1 and PE-A2, the two polymers which have narrow cavities, compared to PE-A1 and PE-P2.

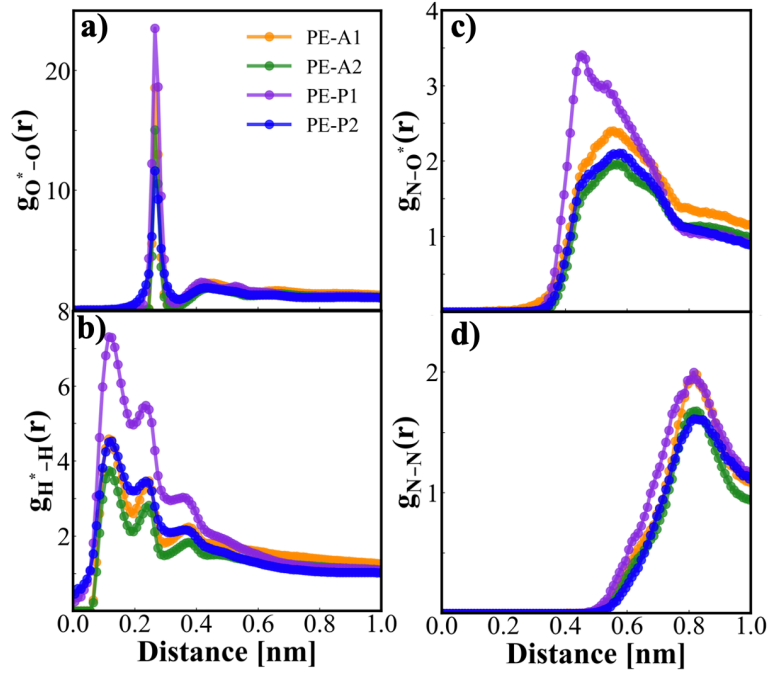

**Figure 5:** Comparison between the radial distribution functions (RDFs) of PE-A1, PE-A2, PE-P1, PE-P2, as labeled. The RDFs shown are for (a) O*—O (hydroxyl oxygen – water oxygen) (b) H*—H (hydroxyl hydrogen – water hydrogen), (c) N—O*(polymer nitrogen – hydroxyl oxygen), and (d) N—N (polymer nitrogen – polymer nitrogen).

The RDFs for the two TFP polymers are given in Figure 6. TFP-25 and TFP-33 have a high hydration number of 30 and 32, respectively. Of the two, we find that TFP-25 has higher $g_{H^*-H}(r)$ and $g_{O^*-O}(r)$ peaks (Figure 6a and 6b), signifying that water is packed closer to hydroxide ions in TFP-25 compared to TFP-33. We also find that $g_{O^*-N}(r)$ is higher in TFP-25, signifying that hydroxide ions are packed tightly to the polymer cation in TFP-25 compared to TFP-33. The cavity size distribution for the two TFP systems are not very different (Figure 3), however, TFP-33 has higher number of cations along the backbone. We also see that the $g_{N-N}(r)$ peak is sharper in TFP-25, suggesting that the polymers chains are more compactly packed in TFP-25 than TFP-33.

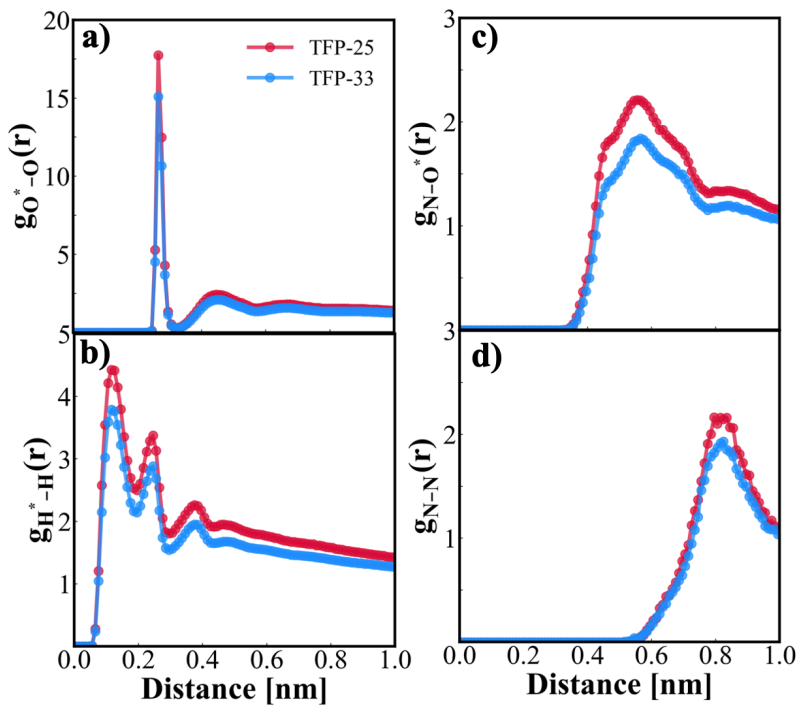

**Figure 6: Comparison between the radial distribution functions (RDFs) of TFP-25 and TFP-33, as labeled. The RDFs shown are for (a) O*—O (hydroxyl oxygen – water oxygen) (b) H*—H (hydroxyl hydrogen – water hydrogen), (c) N—O*(polymer nitrogen – hydroxyl oxygen), and (d) N—N (polymer nitrogen – polymer nitrogen).**

Finally, the RDF for the two PP polymers is shown in Figure 7. The water-hydroxide g(r) peaks are slightly higher in PP-T compared to PP-D, especially $g_{O^*-O}(r)$ (Figure 7b), indicating that the packing of water molecules around the hydroxide ions is slightly greater in PP-T compared to PP-D. The $g_{O^*-N}(r)$ (Figure 7c) peak is markedly higher in PP-T compared to PP-D, revealing that the hydroxide ions are bound tighter to the cationic groups in PP-T compared to PP-D. In line with our observations for the PE polymers, where systems with larger cavities had less intense g(r) peaks indicating that the waters and cations are not closely packed around the hydroxide ions in these polymers, compared to those with smaller cavities, which exhibit sharper g(r) peaks. In addition to the smaller cavities in PP-T, this could also be because of the smaller alkyl chain in PP-T compared to PP-D, which prevents the hydroxide ions from arranging close to the polymer cation. The first peak in $g_{N-N}(r)$ are similar, although PP-T shows a faintly higher first peak, and also has a shoulder at low r compared to PP-D. Overall, we observe that the polymer cavity distribution is inversely correlated with the local packing of different molecules, that is, polymers with small cavities have high g(r) peaks, across the three sets of polymers that we consider. Specifically, the close packing of atoms in PE-P1/PP-T compared to PE-A2/PP-D corresponds to a tighter packing of water molecules around $OH^-$ moieties.

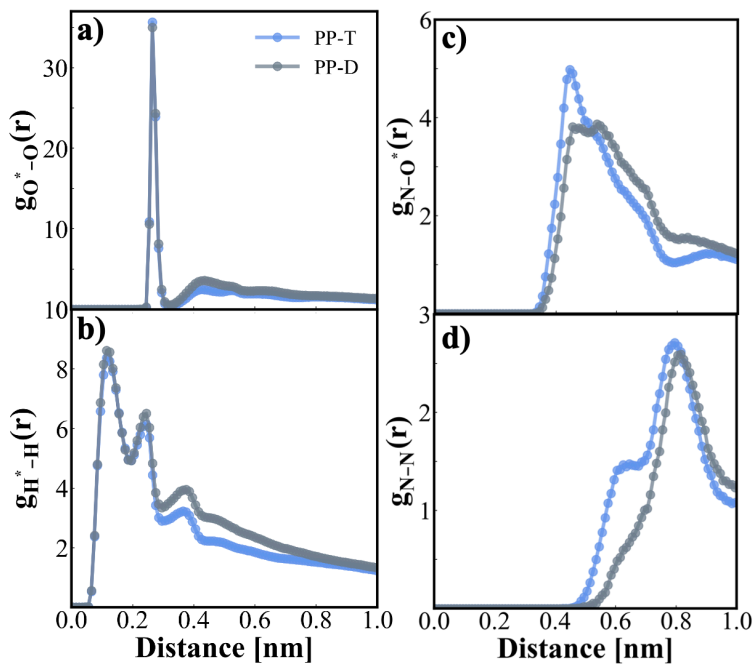

**Figure 7: Comparison between the radial distribution functions (RDFs) of PP-T and PP-D, as labeled. The RDFs shown are for (a) O*—O (hydroxyl oxygen – water oxygen) (b) H*—**

H (hydroxyl hydrogen – water hydrogen), (c) N—O*(polymer nitrogen – hydroxyl oxygen), and (d) N—N (polymer nitrogen – polymer nitrogen).**

We computed the potential of mean force (PMF) from the polymer nitrogen – hydroxyl oxygen ($g_{O^*-N}(r)$) radial distribution function; this data is reported in the Supporting Information, Figure S5. A reduced well depth in the PMF implies that the cation and anion interaction is weak. In such systems, we expect the hydroxide ion to dissociate readily from the functional group, leading to faster transport and conductivity. Of all the systems considered, we find that the well depth in PP-T to be the highest (-4.0 kJ/mol), and that in TFP-33 and PE-A2 to be the lowest (-1.7 kJ/mol).

Details of the membrane microstructure and anion-cation interaction strengths are summarized in Table 2.

**Dynamics**

To explore the influence of chemistry and water channels on the transport and dynamics of hydroxide ions, we compute the MSD as described in the Methods section. To decouple the dynamics of water, polymer and anions, we calculate the MSD of hydroxide ions, water and polymer separately, the results are presented in Figures 8-10. For all polymers, our analysis reveals that at long times, water has the highest mobility, followed by the hydroxide ions, and finally the polymer. At short times however, the hydroxide ions have higher mobility than water. At intermediate times, the hydroxide ions exhibit a plateau due to electrostatic interactions with the polymer cations, which arrests their motion. At long times, the hydroxide ions move in unison with water molecules, specifically, the two MSD slopes are equal, but do not cross each. In this long time regime, vehicular transport of the hydroxide ions facilitated by water molecules, dominates. This aligns with prior research, where it has been shown that the diffusion coefficient of hydroxide ions is significantly lower than that of water molecules at high hydration levels.[46] The motion of the polymer chains in the time scale considered has little to no influence on hydroxide ion transport.

Figure 8 shows the MSD for the four PE systems. Overall, we see the PE-P1 (Figure 8c) and PE-P2 (Figure 8d), with single and double N-methylpiperidinium (NMP) groups, respectively, have low hydroxide MSDs compared to PE-A1 (Figure 8a) and PE-A2 (Figure 8b), which contain trimethylammonium groups. In fact, the hydroxide ions in the PE-P1 system have not reached the diffusive regime (Figure 8c). We also see that in PE-P1 and PE-P2, the intermediate plateau in the hydroxide MSD is more persistent compared to the plateau in the PE-A1 and PE-A2 systems, signifying that the $OH^-$ - $N^+$ interactions are stronger in the former, compared to the latter. This is because the hydroxide ions cannot associate closely with the bulky trimethylammonium groups compared to the N-methylpiperidinium groups. This is also apparent in the potential of mean force (PMF) curves (Figure S4), with PE-P1 having the lowest minima, signifying a tight association of $OH^-$ with NMP groups. Additionally, it can also be seen that water mobility is higher in the PE-A polymers, compared to the PE-P polymers. It is interesting to note that the PE-P1, with the smallest cavity size distribution and strongest g(r) peaks shows the lowest hydroxide mobility, and PE-A2 with the broadest cavity size distribution and the smallest g(r) peaks shows the highest hydroxide mobility. This shows that small cavities lead to tighter associations between the hydroxide ions and polymer cations, as well as restricted water motion, which contribute to an overall decrease in hydroxide dynamics.

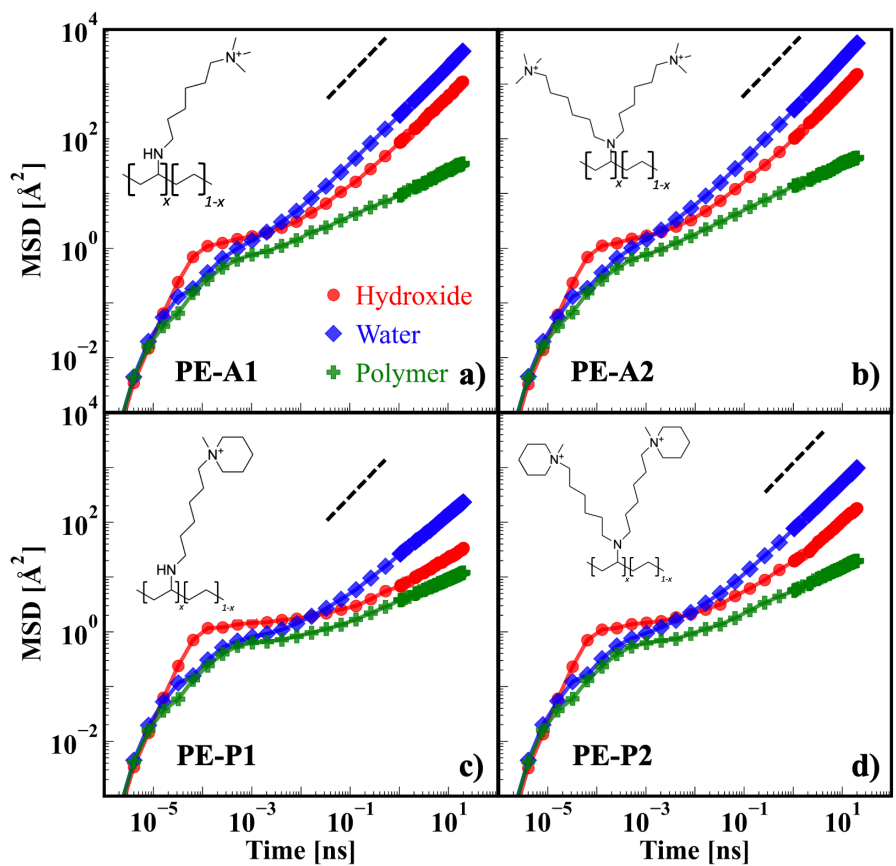

**Figure 8:** Mean squared displacement (MSD) curves for hydroxide ions, waters, and polymer for a) PE-A1 b) PE-A2 c) PE-P1 and d) PE-P2, as labeled. Inset shows monomer chemistry, and the dashed line indicates log-log slope of 1.

The MSD plots for the two TFP polymers are shown in Figure 9, which only vary in their total ion content. We find that the overall trends are consistent with PE polymers. Additionally, both the water and hydroxide dynamics are much higher in TFP compared to PE; we believe this is because of the high hydration numbers of these materials ($\lambda$ = 30 and 32), compared to the PE polymers ($\lambda$ = 18). The plateau in the hydroxide MSD of both TFP polymers is short lived compared to the PE polymers, which is also because of the high number of water molecules in these systems, which enhances vehicular transport.

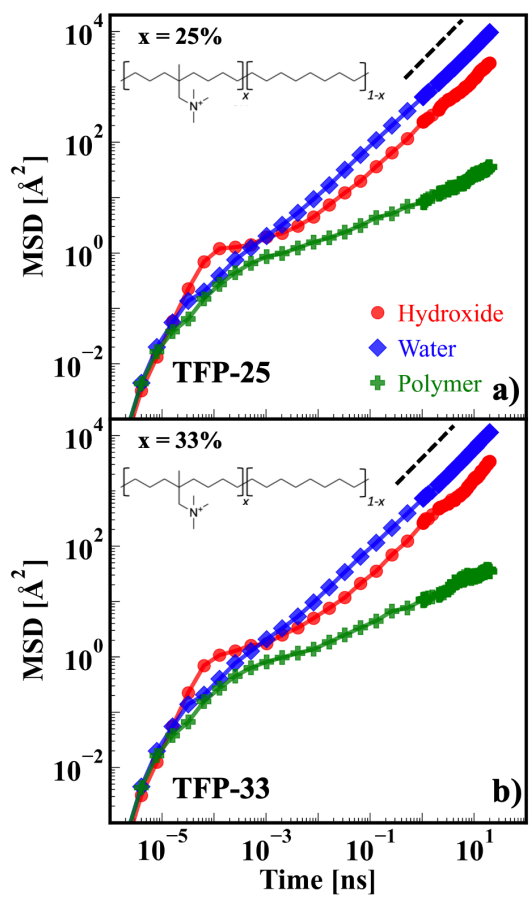

**Figure 9: Mean squared displacement (MSD) curves for hydroxide ions, waters, and polymer for a) TFP-25 and b) TFP-33, as labeled. Inset shows monomer chemistry, and the dashed line indicates log-log slope of 1.**

The MSD curves for the PP polymers are shown in Figure 10. Both these systems have the same hydration number ($\lambda = 12$) but differ in the total number of alkyl chains (Figure 1). Both water and hydroxide dynamics are higher in PP-D compared to PP-T. The subdiffusive plateau for OH$^-$ is more persistent in PP-T compared to PP-D, signifying that the ions take longer to dissociate. The minimum in the PMF is deeper for PP-T compared to PP-D as well (Supporting Information, Figure S4). The dynamics of the three species for both PP polymers is lower compared to the TFP polymers, and somewhat comparable to the PE polymers. Similar to PE-P1, the hydroxide ions in the PP-T have not reached the diffusive regime (Figure 10a) in the timescales considered.

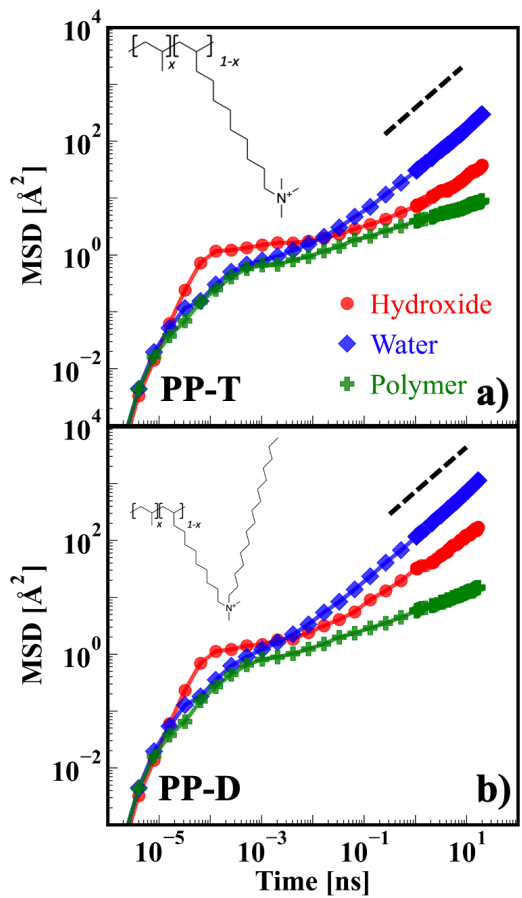

**Figure 10: Mean Squared Displacement (MSD) curves for hydroxide ions, waters, and polymer for the two PP systems averaged over four replicas, as labeled. Inset shows monomer chemistry, and the dashed line indicates log-log slope of 1.**

To quantify the mobility of water molecules and hydroxide ions, we calculate the diffusion coefficient as described in the Methods section. As expected, water has higher diffusion coefficient than the hydroxide in each system. Water and hydroxide diffusion are both highest for the TFP-25 and TFP-33, and lowest for PE-P1 and PP-T. We see a strong correlation between polymer microstructure and hydroxide dynamics, where a more porous microstructure facilitates faster hydroxide transport. We also find that local packing of waters and cations around hydroxide ions is correlated to hydroxide dynamics, as this indicates how effectively the hydroxide ions can dissociate. A clear correlation emerges in Figure 11 between hydroxide and water diffusion across diverse systems. This underscores the significance of water in facilitating hydroxide transport, as elucidated in previous sections. Within each polymer family, we find that our diffusion coefficients are in reasonable agreement with experimental conductivity trends. Specifically, for the systems considered, the measured hydroxide conductivity values for 1) PE polymers are 54, 73, 26, 59 mS/cm for PE-A1, PE-A2, PE-P1, PE-P2, respectively 2) TFP polymers are 40 and 48 mS/cm for TFP-29 and TFP-33, respectively 3) PP polymers are 17 and 19 mS/cm for PP-T and PP-D, respectively. The main differences between the trends in conductivity and diffusion are with the TFP polymers. In our simulations, we find that TFP polymers have the highest $OH^-$ diffusion, however, experimentally, PE-A2 has the highest conductivity. We have chosen to not calculate conductivity in our simulations, as this requires careful treatment of ion-ion correlations and will be the subject of future work. Both simulations and experiments point to the fact that PP polymers have the lowest diffusion and conductivity, respectively. The diffusion coefficients for water and hydroxide ions for the eight systems are reported in Table 2.

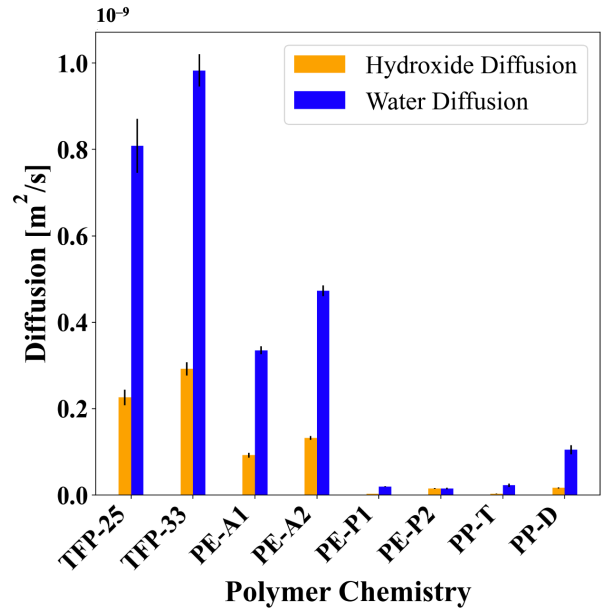

**Figure 11: Diffusion coefficient calculated for eight systems averaged over four independent replicas. Error bars indicate one standard deviation.**

To capture the effect of hydration on ion transport, we consider four hydration levels ($\lambda$=0, 20, 30, 40) in the TFP-25 system, and calculate $OH^-$ diffusion at these hydration levels; this is reported in the Supporting Information, Figure S5. We find that in the dry state, the hydroxide ions do not dissociate from the cations in the time scale of the simulation, and they have very little mobility. As hydration increases, the mobility of $OH^-$ ions increase as well, signifying that water contributes significantly in transporting $OH^-$ ions through the membrane.

**Mechanical Properties**

Lastly, we compute the mechanical properties of these systems, by applying uniaxial tensile deformation, as detailed in the Methods section. The results are presented in Figure 12. To achieve a comparable level of conductivity, AEMs often rely on a higher ionic exchange capacity (IEC). However, this strategy often results in elevated water uptake, as observed in experimental systems. The substantial increase in water absorption, in turn, poses a challenge to the mechanical integrity of the membrane. Overall, we find that the TFP polymers, with the highest hydration levels, have the lowest mechanical toughness (Figure 12b), as seen by the decreased yield stress in these systems (~35 MPa). Amongst the PE polymers, in both experiments[23] and simulations, PE-P1 demonstrates the poorest mechanical performance. Our results show that PE-P2 has the highest yield stress (~50 MPa), likely due to double phenyl groups belonging to the N-methylpiperidinium (Figure 12a). Between the PP polymers, PP-T show significantly higher yield stress compared to PP-D signifying that while the long alkyl enhances conductivity, it lowers the mechanical properties of the membrane. The mechanical properties for the TFP-25 polymer as a function of hydration number was calculated to assess the dependence of mechanical integrity of the membrane with water content, this is reported in the Supporting Information, (Figure S1). When the membrane is dry, the mechanical response is that of a brittle polymer, with a high modulus and yield stress. As the hydration increases, the yield stress and modulus both drop, and the polymer shows decreased mechanical properties.

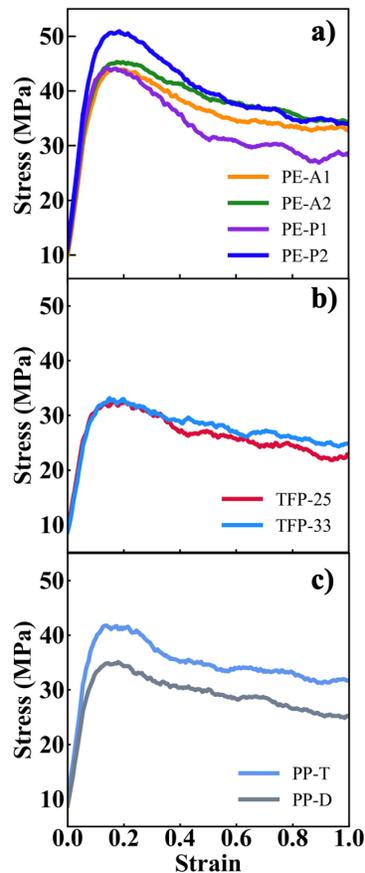

**Figure 12: Stress-strain curves of a) PE, b) TFP, and c) PP polymers undergoing uniaxial tensile deformation, at a strain rate of $10^{-5}$ s$^{-1}$.**

Findings from this study are summarized in Table 2, for the eight systems considered. Overall, there is a strong correlation between the hydration number and hydroxide transport, with PP-T and PE-P1 having the lowest hydration numbers (< 10) and the lowest hydroxide diffusion coefficients. Similarly, TFP polymers with the highest hydration numbers (>30) have the highest hydroxide diffusion. A higher water uptake leads to a greater interconnected void network within the polymer matrix which facilitates the transport of hydroxide ions. The interaction between OH⁻ and the polymer cation are also dependent on the amount of water in the system, as water can screen electrostatic interactions. The mechanical strength is negatively correlated with hydration, with systems with a high hydration number showing reduced yield strength.

**Table 2: Summary of results obtained from MD simulations for the eight systems studied. Two data points reported for average void size and voids/nm$^3$ indicate that the system has a bimodal distribution of pores.**

| System (hydration number) | Average void size (Å) | Voids/ nm$^3$ | OH–N interaction strength (kJ/mol) | OH$^-$ diffusion (m$^2$/sec) | H$_2$O diffusion (m$^2$/sec) | Yield stress (MPa) |
|---|---|---|---|---|---|---|
| PE-A1 (18) | 14 | 1 | -2.3 | 0.100 | 0.30 | 43 |
| PE-A2 (18) | 15 | 2 | -1.7 | 0.150 | 0.450 | 44 |
| PE-P1 (8) | 2 | 0.5 | -3.2 | 0.003 | 0.013 | 43 |
| PE-P2 (11) | 2 | 0.5 | -1.87 | 0.010 | 0.010 | 50 |
| TFP-25 (30) | 2, 15 | 0.4, 3.7 | -2.1 | 0.210 | 0.810 | 32 |
| TFP-33 (32) | 2, 15 | 0.3, 2.5 | -1.7 | 0.270 | 0.970 | 32 |
| PP-T (7) | 2 | 0.6 | -4.0 | 0.004 | 0.012 | 41 |
| PP-D (12) | 2, 15 | 0.7, 0.2 | -3.40 | 0.010 | 0.080 | 33 |

**Conclusions and Outlook**

In this study, we use atomistic MD simulations to understand how the microstructure of polyolefin-based anion exchange membranes influences hydroxide transport, across eight distinct chemistries. We find that a broad cavity size distribution translates to large, interconnected water channels when hydrated, which facilitates faster hydroxide diffusion. Radial distribution function and potentials of mean force reveal the local environment around hydroxide ions, and we see that narrow cavities lead to tighter packing of water molecules around the hydroxide ions, from which the ions do not dissociate easily. System dynamics was investigated by computing the diffusion coefficients, and we find reasonable agreement between calculated diffusivity and experimental conductivity. Mechanical properties of the AEMs were evaluated using uniaxial deformation, and we find that the mechanical integrity of AEMs is inversely impacted by the hydration level. The simulations predict the observed patterns in both conductivity and deformation performance.

While the trends in hydroxide ion transport and conductivity between simulations and experiments are consistent, the conductivity values are not reported for two reasons. The first is that classical MD models used in this work do not account for proton hopping, or Grotthuss transport. Secondly, Nernst-Einstein relation is often utilized to calculate conductivity from the diffusion coefficient, which does not include ion-ion correlations.[47,48] Future work will address these issues by employing approximating reactions using the Reacter forcefield[49] and by employing the Onsager framework[50] to estimate conductivity from simulations. This generic model will allow us to scaleup the screening of novel AEM chemistries via high-throughput simulations.

**Acknowledgement**

This work was supported as part of UNCAGE-ME, an Energy Frontier Research Center funded by the U.S. Department of Energy, Office of Science, Basic Energy Sciences under Award No. DE-SC0012577. The authors acknowledge University of Florida Research Computing for managing the computational resources that have contributed to the research results reported in this publication. http://researchcomputing.ufl.edu.

**Data Availability**

Data files containing force field parameters and structures can be found here - https://github.com/UFSRG/published-work/tree/main/2024_Polyolefin_AEM

**Bibliography**

(1)     Wee, J.-H. Applications of Proton Exchange Membrane Fuel Cell Systems. *Renew. Sustain. Energy Rev* **2007**, *11* (8), 1720–1738. DOI: 10.1016/j.rser.2006.01.005.

(2)     Lu, S.; Pan, J.; Huang, A.; Zhuang, L.; Lu, J. Alkaline Polymer Electrolyte Fuel Cells Completely Free from Noble Metal Catalysts. *Proceedings of the National Academy of Sciences* **2008**, *105* (52), 20611–20614. DOI: 10.1073/pnas.0810041106.

(3)     Vedarajan, R.; Balaji, R.; Ramya, K. Anion Exchange Membrane Fuel Cell: New Insights and Advancements. *WIREs Energy Environ.* **2022**. DOI: 10.1002/wene.466.

(4)     Pan, Z. F.; An, L.; Zhao, T. S.; Tang, Z. K. Advances and Challenges in Alkaline Anion Exchange Membrane Fuel Cells. *Prog. Energy Combust. Sci.* **2018**, *66*, 141–175. DOI: 10.1016/j.pecs.2018.01.001.

(5)     Dong, D.; Wei, X.; Hooper, J. B.; Pan, H.; Bedrov, D. Role of Cationic Groups on Structural and Dynamical Correlations in Hydrated Quaternary Ammonium-Functionalized Poly(p-Phenylene Oxide)-Based Anion Exchange Membranes. *Phys. Chem. Chem. Phys.* **2018**, *20* (29), 19350–19362. DOI: 10.1039/c8cp02211a.

(6)     Wang, J.; Zhao, Z.; Gong, F.; Li, S.; Zhang, S. Synthesis of Soluble Poly(Arylene Ether Sulfone) Ionomers with Pendant Quaternary Ammonium Groups for Anion Exchange Membranes. *Macromolecules* **2009**, *42* (22), 8711–8717. DOI: 10.1021/ma901606z.

(7)     Varcoe, J. R.; Slade, R. C. T. Prospects for Alkaline Anion-Exchange Membranes in Low Temperature Fuel Cells. *Fuel Cells* **2005**, *5* (2), 187–200. DOI: 10.1002/fuce.200400045.

(8)     Ran, J.; Wu, L.; He, Y.; Yang, Z.; Wang, Y.; Jiang, C.; Ge, L.; Bakangura, E.; Xu, T. Ion Exchange Membranes: New Developments and Applications. *J. Memb. Sci.* **2017**, *522*, 267–291. DOI: 10.1016/j.memsci.2016.09.033.

(9)     Lin, B.; Qiu, L.; Qiu, B.; Peng, Y.; Yan, F. A Soluble and Conductive Polyfluorene Ionomer with Pendant Imidazolium Groups for Alkaline Fuel Cell Applications. *Macromolecules* **2011**, *44* (24), 9642–9649. DOI: 10.1021/ma202159d.

(10)    Dang, H.-S.; Weiber, E. A.; Jannasch, P. Poly(Phenylene Oxide) Functionalized with Quaternary Ammonium Groups via Flexible Alkyl Spacers for High-Performance Anion Exchange Membranes. *J. Mater. Chem. A* **2015**, *3* (10), 5280–5284. DOI: 10.1039/C5TA00350D.

(11)    Karibayev, M.; Kalybekkyzy, S.; Wang, Y.; Mentbayeva, A. Molecular Modeling in Anion Exchange Membrane Research: A Brief Review of Recent Applications. *Molecules* **2022**, *27* (11). DOI: 10.3390/molecules27113574.

(12) Ramírez, S. C.; Paz, R. R. Hydroxide Transport in Anion-Exchange Membranes for Alkaline Fuel Cells. In *New trends in ion exchange studies*; Karakuş, S., Ed.; InTech, 2018. DOI: 10.5772/intechopen.77148.

(13) Tomasino, E.; Mukherjee, B.; Ataollahi, N.; Scardi, P. Water Uptake in an Anion Exchange Membrane Based on Polyamine: A First-Principles Study. *J. Phys. Chem. B* **2022**, *126* (38), 7418–7428. DOI: 10.1021/acs.jpcb.2c04115.

(14) Zelovich, T.; Tuckerman, M. E. OH- and H3O+ Diffusion in Model AEMs and PEMs at Low Hydration: Insights from Ab Initio Molecular Dynamics. *Membranes (Basel)* **2021**, *11* (5). DOI: 10.3390/membranes11050355.

(15) Foglia, F.; Berrod, Q.; Clancy, A. J.; Smith, K.; Gebel, G.; Sakai, V. G.; Appel, M.; Zanotti, J.-M.; Tyagi, M.; Mahmoudi, N.; Miller, T. S.; Varcoe, J. R.; Periasamy, A. P.; Brett, D. J. L.; Shearing, P. R.; Lyonnard, S.; McMillan, P. F. Disentangling Water, Ion and Polymer Dynamics in an Anion Exchange Membrane. *Nat. Mater.* **2022**, *21* (5), 555–563. DOI: 10.1038/s41563-022-01197-2.

(16) Zelovich, T.; Vogt-Maranto, L.; Hickner, M. A.; Paddison, S. J.; Bae, C.; Dekel, D. R.; Tuckerman, M. E. Hydroxide Ion Diffusion in Anion-Exchange Membranes at Low Hydration: Insights from Ab Initio Molecular Dynamics. *Chem. Mater.* **2019**, *31* (15), 5778–5787. DOI: 10.1021/acs.chemmater.9b01824.

(17) Tuckerman, M. E.; Chandra, A.; Marx, D. A Statistical Mechanical Theory of Proton Transport Kinetics in Hydrogen-Bonded Networks Based on Population Correlation Functions with Applications to Acids and Bases. *J. Chem. Phys.* **2010**, *133* (12), 124108. DOI: 10.1063/1.3474625.

(18) Tuckerman, M.; Laasonen, K.; Sprik, M.; Parrinello, M. Ab Initio Molecular Dynamics Simulation of the Solvation and Transport of H3O+ and OH- Ions in Water. *J. Phys. Chem.* **1995**, *99* (16), 5749–5752. DOI: 10.1021/j100016a003.

(19) Zhang, W.; van Duin, A. C. T. ReaxFF Reactive Molecular Dynamics Simulation of Functionalized Poly(Phenylene Oxide) Anion Exchange Membrane. *J. Phys. Chem. C* **2015**, *119* (49), 27727–27736. DOI: 10.1021/acs.jpcc.5b07271.

(20) Chen, C.; Tse, Y.-L. S.; Lindberg, G. E.; Knight, C.; Voth, G. A. Hydroxide Solvation and Transport in Anion Exchange Membranes. *J. Am. Chem. Soc.* **2016**, *138* (3), 991–1000. DOI: 10.1021/jacs.5b11951.

(21) Dong, D.; Zhang, W.; van Duin, A. C. T.; Bedrov, D. Grotthuss versus Vehicular Transport of Hydroxide in Anion-Exchange Membranes: Insight from Combined Reactive and Nonreactive Molecular Simulations. *J. Phys. Chem. Lett.* **2018**, *9* (4), 825–829. DOI: 10.1021/acs.jpclett.8b00004.

(22)   Zhang, W.; Dong, D.; Bedrov, D.; van Duin, A. C. T. Hydroxide Transport and Chemical Degradation in Anion Exchange Membranes: A Combined Reactive and Non-Reactive Molecular Simulation Study. *J. Mater. Chem. A* **2019**, *7* (10), 5442–5452. DOI: 10.1039/C8TA10651G.

(23)   Wang, T.; Zhang, Y.; Wang, Y.; You, W. Transition-Metal-Free Preparation of Polyethylene-Based Anion Exchange Membranes from Commercial EVA. *Polymer* **2022**, *262*, 125439. DOI: 10.1016/j.polymer.2022.125439.

(24)   Kostalik, H. A.; Clark, T. J.; Robertson, N. J.; Mutolo, P. F.; Longo, J. M.; Abruña, H. D.; Coates, G. W. Solvent Processable Tetraalkylammonium-Functionalized Polyethylene for Use as an Alkaline Anion Exchange Membrane. *Macromolecules* **2010**, *43* (17), 7147–7150. DOI: 10.1021/ma101172a.

(25)   Zhang, M.; Liu, J.; Wang, Y.; An, L.; Guiver, M. D.; Li, N. Highly Stable Anion Exchange Membranes Based on Quaternized Polypropylene. *J. Mater. Chem. A* **2015**, *3* (23), 12284–12296. DOI: 10.1039/C5TA01420D.

(26)   Hanwell, M. D.; Curtis, D. E.; Lonie, D. C.; Vandermeersch, T.; Zurek, E.; Hutchison, G. R. Avogadro: An Advanced Semantic Chemical Editor, Visualization, and Analysis Platform. *J. Cheminform.* **2012**, *4* (1), 17. DOI: 10.1186/1758-2946-4-17.

(27)   Bannwarth, C.; Ehlert, S.; Grimme, S. GFN2-XTB-An Accurate and Broadly Parametrized Self-Consistent Tight-Binding Quantum Chemical Method with Multipole Electrostatics and Density-Dependent Dispersion Contributions. *J. Chem. Theory Comput.* **2019**, *15* (3), 1652–1671. DOI: 10.1021/acs.jctc.8b01176.

(28)   Wang, J.; Wang, W.; Kollman, P. A.; Case, D. A. Automatic Atom Type and Bond Type Perception in Molecular Mechanical Calculations. *J. Mol. Graph. Model.* **2006**, *25* (2), 247–260. DOI: 10.1016/j.jmgm.2005.12.005.

(29)   Wang, J.; Wolf, R. M.; Caldwell, J. W.; Kollman, P. A.; Case, D. A. Development and Testing of a General Amber Force Field. *J. Comput. Chem.* **2004**, *25* (9), 1157–1174. DOI: 10.1002/jcc.20035.

(30)   Jakalian, A.; Jack, D. B.; Bayly, C. I. Fast, Efficient Generation of High-Quality Atomic Charges. AM1-BCC Model: II. Parameterization and Validation. *J. Comput. Chem.* **2002**, *23* (16), 1623–1641. DOI: 10.1002/jcc.10128.

(31)   Zhang, X. B.; Liu, Q. L.; Zhu, A. M. An Improved Fully Flexible Fixed-Point Charges Model for Water from Ambient to Supercritical Condition. *Fluid Phase Equilib.* **2007**, *262* (1–2), 210–216. DOI: 10.1016/j.fluid.2007.09.005.

(32)   Fortunato, M. E.; Colina, C. M. Pysimm: A Python Package for Simulation of Molecular Systems. *SoftwareX* **2017**, *6*, 7–12. DOI: 10.1016/j.softx.2016.12.002.

(33)     Martínez, L.; Andrade, R.; Birgin, E. G.; Martínez, J. M. PACKMOL: A Package for Building Initial Configurations for Molecular Dynamics Simulations. *J. Comput. Chem.* **2009**, *30* (13), 2157–2164. DOI: 10.1002/jcc.21224.

(34)     Zheng, Y.; Ash, U.; Pandey, R. P.; Ozioko, A. G.; Ponce-González, J.; Handl, M.; Weissbach, T.; Varcoe, J. R.; Holdcroft, S.; Liberatore, M. W.; Hiesgen, R.; Dekel, D. R. Water Uptake Study of Anion Exchange Membranes. *Macromolecules* **2018**, *51* (9), 3264–3278. DOI: 10.1021/acs.macromol.8b00034.

(35)     Pasquini, L.; Di Vona, M. L.; Knauth, P. Effects of Anion Substitution on Hydration, Ionic Conductivity and Mechanical Properties of Anion-Exchange Membranes. *New J. Chem.* **2016**, *40* (4), 3671–3676. DOI: 10.1039/C5NJ03212A.

(36)     Hagesteijn, K. F. L.; Jiang, S.; Ladewig, B. P. A Review of the Synthesis and Characterization of Anion Exchange Membranes. *J. Mater. Sci.* **2018**, *53* (16), 11131–11150. DOI: 10.1007/s10853-018-2409-y.

(37)     Dekel, D. R. Review of Cell Performance in Anion Exchange Membrane Fuel Cells. *J. Power Sources* **2017**, *375*, 158–169. DOI: 10.1016/j.jpowsour.2017.07.117.

(38)     Larsen, G. S.; Lin, P.; Hart, K. E.; Colina, C. M. Molecular Simulations of PIM-1-like Polymers of Intrinsic Microporosity. *Macromolecules* **2011**, *44* (17), 6944–6951. DOI: 10.1021/ma200345v.

(39)     Thompson, A. P.; Aktulga, H. M.; Berger, R.; Bolintineanu, D. S.; Brown, W. M.; Crozier, P. S.; in 't Veld, P. J.; Kohlmeyer, A.; Moore, S. G.; Nguyen, T. D.; Shan, R.; Stevens, M. J.; Tranchida, J.; Trott, C.; Plimpton, S. J. LAMMPS - a Flexible Simulation Tool for Particle-Based Materials Modeling at the Atomic, Meso, and Continuum Scales. *Comput. Phys. Commun.* **2022**, *271*, 108171. DOI: 10.1016/j.cpc.2021.108171.

(40)     Willmore, F. T. A Toolkit for the Analysis and Visualization of Free Volume in Materials. In *Proceedings of the 1st Conference of the Extreme Science and Engineering Discovery Environment on Bridging from the eXtreme to the campus and beyond - XSEDE '12*; ACM Press: New York, New York, USA, 2012; p 1. DOI: 10.1145/2335755.2335826.

(41)     Jiang, Y.; Willmore, F. T.; Sanders, D.; Smith, Z. P.; Ribeiro, C. P.; Doherty, C. M.; Thornton, A.; Hill, A. J.; Freeman, B. D.; Sanchez, I. C. Cavity Size, Sorption and Transport Characteristics of Thermally Rearranged (TR) Polymers. *Polymer* **2011**, *52* (10), 2244–2254. DOI: 10.1016/j.polymer.2011.02.035.

(42)     Willmore, F. T.; Wang, X.; Sanchez, I. C. Free Volume Properties of Model Fluids and Polymers: Shape and Connectivity. *J. Polym. Sci. B Polym. Phys.* **2006**, *44* (9), 1385–1393. DOI: 10.1002/polb.20793.

(43)   Al Otmi, M. A.; Willmore, F.; Sampath, J. Structure, Dynamics, and Hydrogen Transport in Amorphous Polymers: An Analysis of the Interplay between Free Volume Element Distribution and Local Segmental Dynamics from Molecular Dynamics Simulations. *Macromolecules* **2023**. DOI: 10.1021/acs.macromol.3c01508.

(44)   Wernisch, B. M.; Al Otmi, M.; Beauvais, E. J.; Sampath, J. Evolution of Free Volume Elements in Amorphous Polymers Undergoing Uniaxial Deformation: A Molecular Dynamics Simulations Study. *Mol. Syst. Des. Eng.* **2024**. DOI: 10.1039/D3ME00148B.

(45)   Hossain, M. D.; Reid, J. C.; Lu, D.; Jia, Z.; Searles, D. J.; Monteiro, M. J. Influence of Constraints within a Cyclic Polymer on Solution Properties. *Biomacromolecules* **2018**, *19* (2), 616–625. DOI: 10.1021/acs.biomac.7b01690.

(46)   Dubey, V.; Daschakraborty, S. Translational Jump-Diffusion of Hydroxide Ion in Anion Exchange Membrane: Deciphering the Nature of Vehicular Diffusion. *J. Phys. Chem. B* **2022**, *126* (12), 2430–2440. DOI: 10.1021/acs.jpcb.2c00240.

(47)   Marioni, N.; Zhang, Z.; Zofchak, E. S.; Sachar, H. S.; Kadulkar, S.; Freeman, B. D.; Ganesan, V. Impact of Ion-Ion Correlated Motion on Salt Transport in Solvated Ion Exchange Membranes. *ACS Macro Lett.* **2022**, *11* (11), 1258–1264. DOI: 10.1021/acsmacrolett.2c00361.

(48)   France-Lanord, A.; Grossman, J. C. Correlations from Ion Pairing and the Nernst-Einstein Equation. *Phys. Rev. Lett.* **2019**, *122* (13), 136001. DOI: 10.1103/PhysRevLett.122.136001.

(49)   Gissinger, J. R.; Jensen, B. D.; Wise, K. E. REACTER: A Heuristic Method for Reactive Molecular Dynamics. *Macromolecules* **2020**, *53* (22), 9953–9961. DOI: 10.1021/acs.macromol.0c02012.

(50)   Fong, K. D.; Self, J.; McCloskey, B. D.; Persson, K. A. Onsager Transport Coefficients and Transference Numbers in Polyelectrolyte Solutions and Polymerized Ionic Liquids. *Macromolecules* **2020**, *53* (21), 9503–9512. DOI: 10.1021/acs.macromol.0c02001.

Supplementary Information

for

**Hydroxide Transport and Mechanical Properties of Polyolefin-Based Anion Exchange Membranes from Atomistic Molecular Dynamics Simulations**

Mohammed Al Otmi,[1] Ping Lin,[2] William Schertzer,[3] Coray M. Colina,[2,3] Rampi Ramprasad,[3] Janani Sampath[1,*]

[1]Department of Chemical Engineering, University of Florida, Gainesville, Florida, 32611

[2]Department of Chemistry, University of Florida, Gainesville, Florida, 32611

[3]Department of Materials Science, University of Florida, Gainesville, Florida, 32611

[4]School of Materials Science and Engineering, Georgia Institute of Technology, Atlanta, Georgia, 30332

[*]jsampath@ufl.edu

**Qualitative Analysis of Water Channels**

Three figures (S1, S2, S3) are provided to show the water channels in the chemistries. Specifically, Figure S1 displays representative snapshots unveiling the water channels within four PE systems. Figure S2 shows snapshots illustrating the water channels in TFP-25 and TFP-33. Figure S3 shows representative snapshots depicting the water channels within two PP systems.

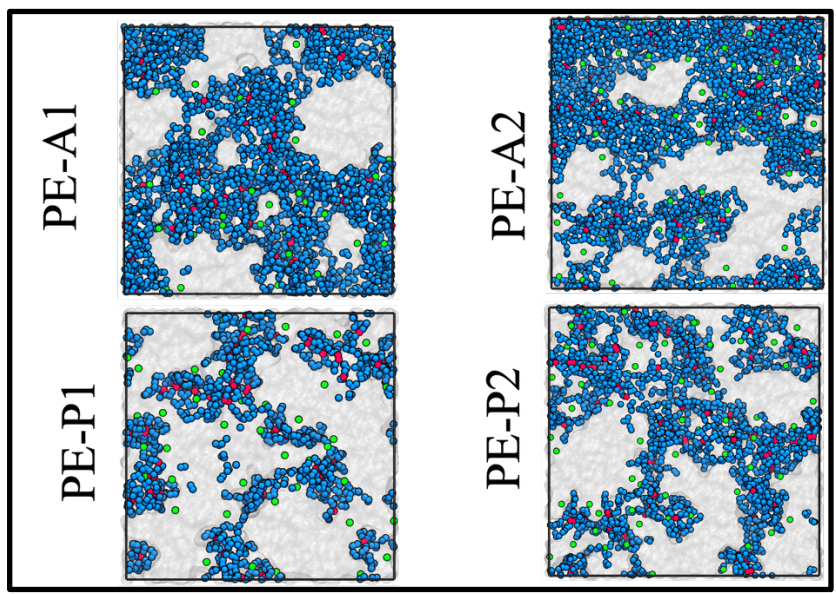

**Figure S1: Representative snapshots showing the water channels in the four PE-13 systems, as labeled. Water molecules are in blue, hydroxide ions are in red, nitrogen atoms are in green, and the other polymer groups are transparent and gray.**

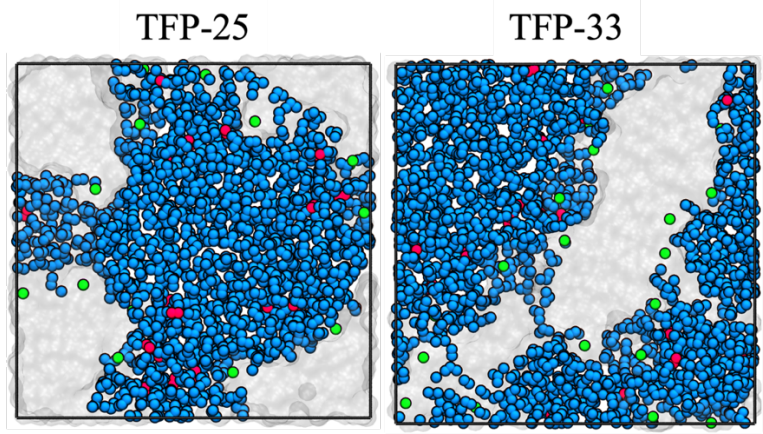

**Figure S2:** Representative snapshots showing the water channels in the two PE systems, as labeled. Water molecules are in blue, hydroxide ions are in red, nitrogen atoms are in green, and the other polymer groups are transparent and gray.

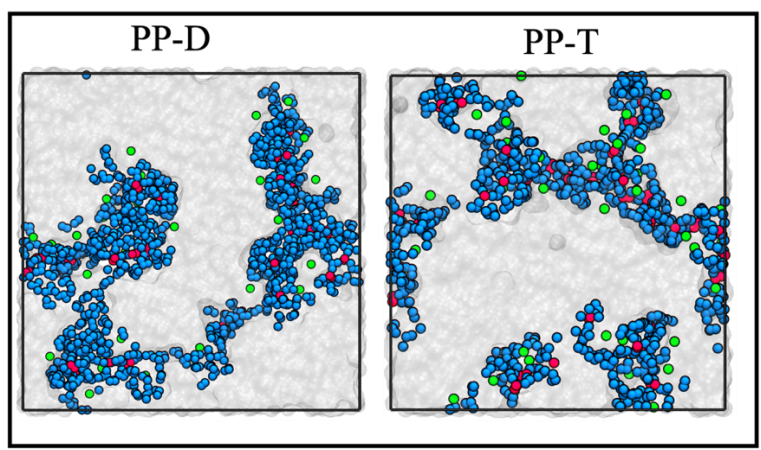

**Figure S3:** Representative snapshots showing the water channels in the two PP-20 systems, as labeled. Water molecules are in blue, hydroxide ions are in red, nitrogen atoms are in green, and the other polymer groups are transparent and gray.

**Potential Mean Field (PMF)**

The potential mean field (PMF) profiles are derived from the radial distribution function of N-O*, where N represents nitrogen situated on the cationic site and O* denotes oxygen within the hydroxide ion. The calculation of the potential mean field is determined by the following equation.

$$\mathbf{PMF = -RT \ln(g(r))}$$

Here, R denotes the gas constant, T represents the temperature in Kelvin, and $g(r)$ is the N-O* radial distribution function. The PMF serves as a measure to assess the electrostatic interactions between hydroxide ions and cationic functional groups. The PMF profiles for the eight distinct systems under discussion are depicted in Figure S4. The depth of the well correlates with the strength of electrostatic interactions. Systems exhibiting reduced well depth, like PE-A2, demonstrate weaker electrostatic interactions between the ammonium and hydroxide anions, resulting in lower dissociation energy and faster OH dynamics. Conversely, systems with a deep PMF well, such as PE-P1, experience stronger interactions and consequently dissociate less readily, leading to slower OH dynamics.

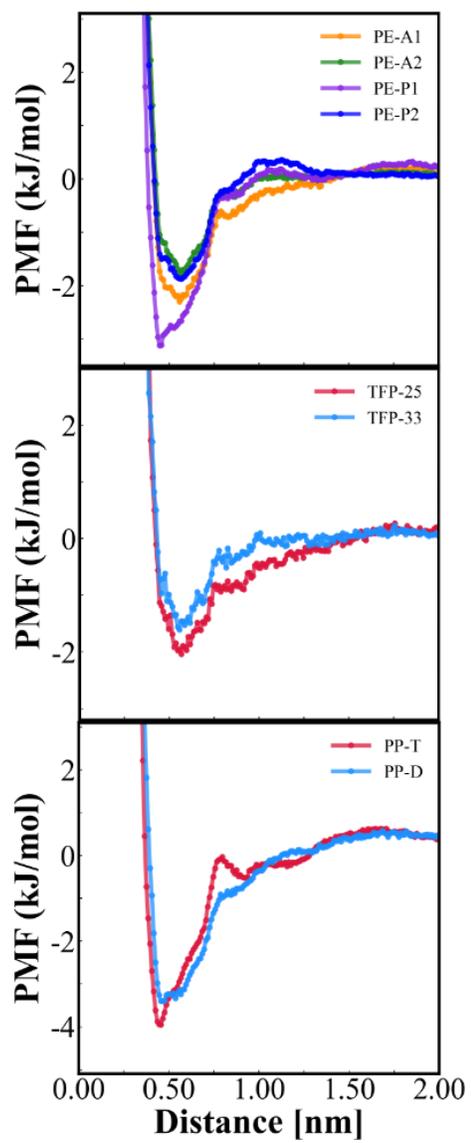

**Figure S4: Potential Mean Field (PMF) profile as a function of the N-O distance where N is the cationic site, and the O\* is the oxygen in the hydroxide ion.**

**Impact of Water Uptake**

The conductivity and mechanical performance of the systems are intricately linked to the water uptake (hydration number). In the previous analyses, the experimental water uptake data from prior studies [1–3] served as the basis for our investigation. Under different water uptake values, both conductivity and mechanical properties change significantly. Higher water uptake usually leads to formation of larger water channels and a significant increase in the vehicular diffusion of hydroxide. Simultaneously, the presence of water weakens the structural robustness and leads to a decline in the material mechanical properties. To elucidate the influence of water, we analyze the change in conductivity and mechanical performance while systematically varying the hydration number for TFP-25. Figure S5 depicts the increase of hydroxide MSD and, consequently, diffusion with increasing hydration numbers. At the same time, the stress-strain reveals a decrease in mechanical properties as the hydration number is elevated.

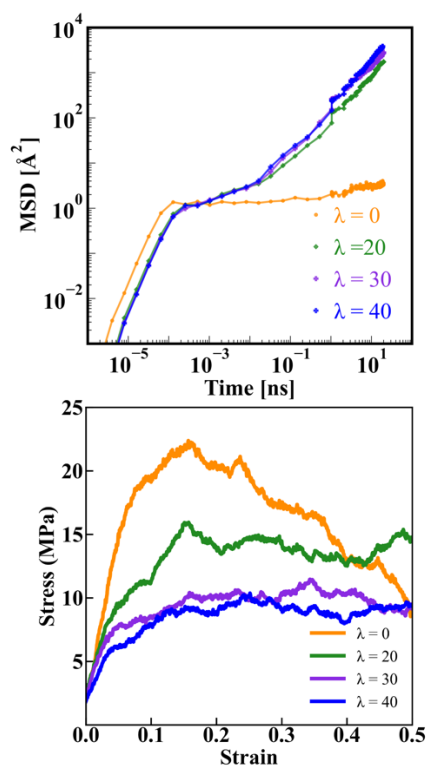

**Figure S5: Impact of hydration number on the conductivity and mechanical performance of the membrane module. Hydration Number ($\lambda$) ranges from 0 - 40 water molecules per $OH^-$.**

# References

(1) Wang, T.; Zhang, Y.; Wang, Y.; You, W. Transition-Metal-Free Preparation of Polyethylene-Based Anion Exchange Membranes from Commercial EVA. *Polymer* **2022**, *262*, 125439. DOI: 10.1016/j.polymer.2022.125439.

(2) Kostalik, H. A.; Clark, T. J.; Robertson, N. J.; Mutolo, P. F.; Longo, J. M.; Abruña, H. D.; Coates, G. W. Solvent Processable Tetraalkylammonium-Functionalized Polyethylene for Use as an Alkaline Anion Exchange Membrane. *Macromolecules* **2010**, *43* (17), 7147–7150. DOI: 10.1021/ma101172a.

(3) Zhang, M.; Liu, J.; Wang, Y.; An, L.; Guiver, M. D.; Li, N. Highly Stable Anion Exchange Membranes Based on Quaternized Polypropylene. *J. Mater. Chem. A* **2015**, *3* (23), 12284–12296. DOI: 10.1039/C5TA01420D.